\documentclass[12pt,a4paper]{article}

\makeatletter

\renewcommand\section{\@startsection%{section}{1}{\z@}
 {section}{1}{0pt}%name, level, indent
  {-\baselineskip}%             beforeskip
  {0.5\baselineskip}%            afterskip
{\raggedright\reset@font\normalsize\bfseries\mathversion{bold}}}%style
\renewcommand{\subsection}{\@startsection%{subsection}{2}{\z@}
{subsection}{2}{0mm}%name, level, indent
  {-\baselineskip}%             beforeskip
{0.5\baselineskip}%            afterskip
{\normalfont\normalsize\itshape}}
\renewcommand{\subsubsection}{\@startsection%{subsection}{2}{\z@}
{subsubsection}{3}{0mm}%name, level, indent
  {-\baselineskip}%             beforeskip
  {0.3\baselineskip}%            afterskip
{\normalfont\normalsize\itshape}}

\long\def\@makefntext#1{%
  \noindent \hb@xt@ 0.2em{\hss \@makefnmark}\hskip0.5em\relax#1}%

\makeatother

\usepackage{lmodern}
\usepackage{amsmath,amssymb,amsthm}
\usepackage[USenglish,british,american,australian]{babel}
\usepackage{wrapfig}
\usepackage{enumerate}
\usepackage{graphicx,color}
\usepackage{float}
\usepackage{natbib}
\bibpunct[:]{(}{)}{,}{a}{}{,}
\usepackage{setspace}
\usepackage[margin=1in]{geometry}
\usepackage{soul}
\usepackage[title]{appendix}
\newcommand{\E}{\mathbb{E}}
\newcommand{\Var}{\mathrm{Var}}

\theoremstyle{plain}
\newtheorem{theorem}{Theorem}
\newtheorem{corollary}{Corollary}
\newtheorem{lemma}{Lemma}
\newtheorem{proposition}{Proposition}

\theoremstyle{definition}
\newtheorem{definition}{Definition}

\newtheorem{remark}{Remark}
\newtheorem{assumption}{Assumption}

\newcounter{num}

\begin{document}
\bibliographystyle{plainnat}

\title{Insuring uninsurable income\thanks{The earlier version of this paper was circulated under the title ``Mutual insurance for uninsurable income". }
}
\author{Michiko Ogaku\footnote{Corresponding author. Faculty of Economics, Nagasaki University, 4-2-1 Katafuchi, Nagasaki, 850-8506 Japan, e-mail: ohgaku@nagasaki-u.ac.jp}
}
\date{17 March 2026}
\maketitle%\thispagestyle{empty}

%\newpage

\begin{abstract}
We study dynamic mechanism design in a pure-exchange economy with privately observed idiosyncratic income. 
In the standard infinitely lived hidden-income benchmark of Green (1987) and Thomas--Worrall (1990), constrained-efficient allocations exhibit immiseration. 
We propose a simple recursive mechanism---adapted from Marcet--Marimon (1992)---that shifts each income shock forward by one period, keeps promised utilities in a bounded set, 
and, under a transparent ``moderate risk-aversion'' condition, delivers sequential efficiency. In a stationary \emph{overlapping-generations} setting, we further show that under additional symmetry and curvature assumptions, a second-order approximation yields a sufficient condition for period-by-period budget balance; early cohorts pre-fund later transfers; for suitable initial promises, all cohorts are better off than under autarky. 
Our analysis uses a single state (promised utility), closed-form transfers, and a Bellman verification.
\end{abstract}
%\noindent

\vspace{6pt}


\newpage
\section{Introduction}
Private information can render idiosyncratic income effectively uninsurable. Since the seminal work of \cite{Green-1987} and 
\cite{thomas-warrall-1990}, the constrained-efficient hidden-income contract trades insurance off against incentives and features a drift of average promised utilities toward a lower bound---\emph{immiseration}---with cross-sectional dispersion rising over time.\footnote{Earlier papers sometimes use \emph{immiserization}\citep[e.g.,][]{Phelan-1998, zhang2009dynamic}. }
 When agents are almost perfectly patient, the drift can vanish \citep{Carrasco-Fuchs-Fukuda-2019}, but recent work with persistent private information, \cite{bloedel2025insurance}, 
 shows that immiseration remains robust beyond the i.i.d. benchmark. 

A different route was proposed by \cite{Marcet-Marimon-1992}. Allowing agents to make productivity-enhancing investments, they designed a transfer rule that shifts risk across periods for the same individual. Capital accumulation enlarges the future consumption set and breaks the immiseration trap; the associated recursive saddle-point representation, later formalized by \cite{Marcet-Marimon-2019} makes the analysis more tractable. Yet the mechanism has not been explored in a pure-exchange economy without investment or storage, where the scope for avoiding downward drift is a priori unclear. 

This paper adapts the Marcet--Marimon risk-shifting mechanism, that is,  the current private shock changes tomorrow’s promised utility rather than today’s consumption, to an exchange economy in which agents cannot save. Promised utility is the sole state variable, so the model yields a closed-form characterization. 

Relative to \cite{Green-1987} and \cite{thomas-warrall-1990}, this paper changes three ingredients: agents have stochastic finite lifetimes in a stationary OLG environment, the mechanism is restricted to full within-period insurance, and incentives are provided through continuation value alone. The paper studies the consequences of that restricted class of mechanisms. 

The contribution of this paper is threefold. (i) A recursive, one-state dynamic mechanism that postpones shocks one period ahead and eliminates immiseration while preserving sequential incentive compatibility; (ii) a transparent sufficient condition for sequential efficiency under moderate risk aversion; (iii) in a stationary overlapping-generations society, an explicit inequality on the initial promise that suffices for budget sustainability, with a simple economic interpretation; early cohorts pre-fund later transfers; the mechanism therefore delivers higher expected lifetime utility than autarky for all cohorts for suitable initial promises. 

Unlike \cite{fernandes2000recursive}, who enlarge the recursive state because 
serial correlation in private shocks destroys the one-dimensional promised-utility 
representation, this paper retains an i.i.d. hidden-income environment but identifies 
a mechanism in which current private shocks are absorbed by continuation promises. 
Under Assumption \ref{ass:moderate-risk}, this yields a simple one-state Bellman representation with full within-period insurance and bounded promised utility. 

\cite{farhi-Werning_2007} and \cite{phelan_2006} reinterpret and extend the classic
hidden-information benchmark by replacing an infinitely lived individual with a dynasty
of altruistically linked one-period-lived agents. In that dynastic literature, the central
issue is how welfare, or promised utility, is inherited across parent and child under the
incentive logic of Green (1987) and Thomas--Worrall (1990). By contrast, this paper
studies a stationary overlapping-generations environment with stochastic survival.
Its mechanism provides incentives by shifting the current private
shock into next period's promised utility; under Assumption
\ref{ass:moderate-risk}, this is consistent with full within-period insurance.

\cite{demange2002optimality} studies Pareto properties of equilibrium risk sharing in stochastic OLG economies and identifies conditions under which adding a PAYG system at equilibrium is not Pareto-improving. By contrast, this paper studies a planner's sequentially incentive-compatible transfer mechanism under hidden income. The mechanism is not a pure equilibrium PAYG reform: it uses continuation promises and, in the OLG extension, can involve transition pre-funding by early cohorts. Hence, the two papers address different institutional questions and need not deliver the same welfare conclusion.

The organization of the paper is as follows.   
Section \ref{sec:model} sets up the model and transfer rule. Section \ref{sec:mechanism-M} 
proves efficiency and boundedness, and analyzes inter-generational sustainability. Section \ref{sec:numerical_example} presents numerical illustrations. Section \ref{sec:Bellman} verifies optimality via a Bellman equation, and  Section \ref{sec:conclusion} concludes.

\section{Model}\label{sec:model}
\subsection{Environment}
We consider a risk-neutral planner allocating transfers across a continuum of risk-averse individuals indexed by birth cohorts. 
Individuals survive to the next period with probability $\alpha \in (0,1)$ and discount at rate $r \in (0,1)$; their effective discount factor is $\beta=\alpha r$. 
In each period, a newborn cohort enters so that the cross-section is stationary.\footnote{A similar stationary population model is used in \cite{fujiwara2009voluntarily} for their Prisoner's Dilemma game.} Each individual draws an i.i.d. income shock $e_t \in E:=\{e^1, e^2, \dots, e^M\} \subset \mathbb{R}_+$ with $e^i<e^j $ for $i<j \in \{1,\dots, M\}$, $M \geq 2$. 
The planner commits to a recursive transfer mechanism $\mathcal{M}$ mapping the current promised $\lambda_t>0$ into a transfer $\tau_t$ and next period's promise $\lambda_{t+1}$. The single state variable is the promised utility $\overline{v}_1(\lambda)$, which is strictly increasing in $\lambda$. We use $\lambda_t$ as a dual representation of promised utility $w_t=\overline{v}_1(\lambda_t)$; all results could equivalently be written in $w_t$. 

\subsection{Symmetric-information benchmark} 
First, we derive an optimal transfer function which, under symmetric income information, ensures a level of utility corresponding to a promise $\lambda_0>0$ . Following \cite{Marcet-Marimon-1992}, we will use this promise $\lambda_0$ as a state variable under asymmetric income information. For this purpose, in this subsection only, we consider the case where individuals cannot hide their income from the planner in each period.

At period $0$, the planner is considering an efficient transfer mechanism that increases individuals' lifetime utility as much as possible, but it cannot afford an infinite net subsidy.  The planner solves the following problem, which is normalized by multiplying $(1-\beta)$, having decided on an upper bound for the net subsidy.
\begin{align*}
(\ast)%\begin{cases}
\quad \max_{\{\tau_t\}_{t=0}^{\infty} } (1-\beta ) \E_0 \left[\sum_{t=0}^{\infty} \beta^t u(c_t)\right]
%\end{align*}
\quad \text{s.t. } 
%\begin{align*}
\quad (1-\beta )\E_0 \left[\sum_{t=0}^{\infty} \beta^t( \tau_t )\right]\leq \overline{C},
%\end{cases}
\end{align*}
where $\overline{C}>0$ is an exogenous upper bound on the present value of net subsidies, 
$\E_0$ is the conditional expectation given the information in period $0$, 
$u: \mathbb{R}_+ \to \mathbb{R}$ is the individual's utility function, $c_t=e_t+\tau_t$ is the consumption in period $t$ starting from the individual's birth period. The utility function $u$ is assumed to be $u'>0$, $u''<0$, to satisfy the Inada conditions ($\lim_{c \to 0} u'(c)=+\infty$, $\lim_{c \to \infty} u'(c)=0$) and to be bounded from above. Boundedness of $u$ is used later in the proof of Theorem \ref{thm:main} to show that the promised-utility martingale is bounded and hence convergent.

Without specifying $\overline{C}$, $\{\tau_t\}_{t=0}^{\infty}$ is determined by $\lambda_0$, and the same $\lambda_0$ determines $\overline{C}$ as well. 
Let $\mu>0$ be the multiplier on the subsidy constraint and put $\lambda_0=1/\mu$. 
The problem $(\ast)$ can be rewritten as the following problem. 
\begin{align}
\max_{\{\tau_t\}_{t=0}^{\infty} } &(1-\beta ) \E_0 \left[\sum_{t=0}^{\infty} \beta^t [\lambda_0 u(c_t)-\tau_t]\right]. \label{eq:PO}
\end{align}
 From the first order condition, 
\begin{align*}
u'(c_t)=\frac{1}{\lambda_0} \text{  for } t \in \mathbb{N}\cup \{0\}
\end{align*}
and the optimal policy delivers a constant consumption
$c^{\ast}(\lambda_0)=(u')^{-1}(\lambda_0^{-1})$ and transfer $\tau^{\ast}(e \vert \lambda_0)=(u')^{-1}(\lambda_0^{-1})-e$, where $\tau^*(\cdot \ \vert \lambda_0)$ is a time-invariant transfer function. We record two objectives for future use: 
Let $v_1: \mathbb{R}_{++} \times \mathbb{R}_+ \to \mathbb{R}$ be a function expressing each individual's lifetime utility provided by the transfer mechanism. Let $v_2: \mathbb{R}_{++} \times \mathbb{R}_+ \to \mathbb{R}$ be a function expressing the expected balance of transfers. Then, for given $\lambda_0>0$ and $e_0 \in E$, these values are written as follows. 
\begin{align*}
v_1(\lambda_0, e_{0})&:= (1-\beta) \E_0[ \sum_{t=0}^{\infty} \beta^t u(c_t)]=u((u')^{-1}(\lambda_0^{-1})) =: \overline{v}_1(\lambda_0) \, \text{(promised utility)}\\
v_2(\lambda_0, e_{0})&:=(1-\beta) \E_{0}[\sum_{t=0}^{\infty}\beta^t (-\tau_t)] \,\text{(expected budget impact)}.
\end{align*} 
\noindent
The map $\lambda \mapsto \overline{v}_1(\lambda)$ is strictly increasing and will serve as the state variable once income becomes private. We also write 
\[
\overline{v}_2(\lambda_t):=\E[(1-\beta) \E_{t}[\sum_{s=t}^{\infty}\beta^{s-t} (-\tau^{\ast} (e_s \vert \lambda_s))]].
\] 
\begin{remark}
(Lifetime value functional)
Let $\mathcal{T}$ be the set of all admissible control functions. 
Given an initial promise $\lambda_t$ and any admissible transfer rule $\tau \in \mathcal{T}$, each individual's discounted lifetime utility is  
\begin{align}
J(\lambda_t, t, \tau) = (1-\beta ) \E_t \left[\sum_{k=t}^{\infty} \beta^{k-t}  u(e_k + \tau(e_k \vert \lambda_k)) \right] \quad \text{for }\tau \in \mathcal{T}. \label{eq:def_of_J}
\end{align}
This function will be used in Section \ref{sec:Bellman} to verify that our mechanism solves the associated Bellman equation. 
\end{remark}
 \subsection{Private information and mechanism design}
If income is private information, the planner has to ask individuals to declare their income. 
In general, a mechanism $\Gamma$ that arranges transfers $\{\tau_t\}_{t=0}^{\infty}$ needs  to satisfy sequential incentive compatibility defined below. 
\begin{definition}[Sequential incentive compatibility]
Given the current promise $\lambda$, truthful reporting maximizes the continuation value. Equivalently, 
\[
(1-\beta)u(e + \tau(e \vert \lambda)) + \beta \overline{v}_1(\lambda'(e \vert \lambda)) 
\geq (1-\beta) u(e + \tau(\tilde{e} \vert \lambda)) + \beta \overline{v}_1(\lambda'(\tilde{e} \vert \lambda)) \quad \forall e, \tilde{e} \in E.
\]
\end{definition}

We further evaluate a mechanism according to the following notion. 
\begin{definition}[Sequentially efficient mechanism]
 Fix a current state $(\lambda, e)$. Let $\mathcal{C}(\lambda, e)$ be the set of feasible sequentially incentive-compatible continuation mechanisms from that state. For any $\Gamma \in \mathcal{C}(\lambda,e)$, let 
\[
V_1^{\Gamma}(\lambda, e)
\] denote the individual's expected discounted continuation utility, and let 
\[
V_2^{\Gamma}(\lambda, e)
\] denote the planner's expected discounted continuation payoff (equivalently, minus the expected discounted transfer cost). 
We say that $\Gamma'$ Pareto dominates $\Gamma$ at $(\lambda, e)$ if 
\[
V_1^{\Gamma'}(\lambda, e) \geq V_1^{\Gamma}(\lambda,e), \quad V_2^{\Gamma'}(\lambda,e) \geq V_2^{\Gamma}(\lambda,e),
\] with at least one strict inequality. 
A mechanism $\Gamma$ is sequentially efficient at $(\lambda, e)$ if it is sequentially incentive compatible and is not Pareto dominated at $(\lambda, e)$ by any other sequentially incentive-compatible continuation mechanism. 
\end{definition}

\begin{remark}
In this paper, Pareto comparisons are made over the continuation payoff pair
$(V_1,V_2)$, conditional on the current state $(\lambda,e)$. Accordingly, for any
fixed state $(\lambda,e)$, any solution to problem \eqref{eq:PO} yields a
Pareto-efficient continuation payoff pair in this sense.
\end{remark}

In addition to taking these things into account, 
since it is known to lead to an immiseration outcome, the planner does not hope to manipulate the transfer function $\tau^*$. The planner takes an alternative approach. It is to vary the promise $\lambda$ in each period. 
This could be an incentive to tell the truth. 
For a given $\lambda$ in a given period, the planner promises to ensure that the lifetime utility level $\overline{v}_1(\lambda)$, and the promise for the next period is renewed taking into account the current promise and the income reported in the current period. If an individual truthfully reports income $e$ in a period $t$, then the lifetime utility from that period should be    
\begin{align}
\overline{v}_1(\lambda)=\E\big[(1-\beta)u(e+\tau^*(e \vert \lambda))+\beta\overline{v}_1(\lambda'(e \vert \lambda))\big], \label{ineq:PK}
\end{align}
where $\lambda'(\cdot \vert \lambda):E  \to \mathbb{R}_{++}$ is the promise for the next period. 
If $\lambda'$ satisfies \eqref{ineq:PK}, then $\E[\overline{v}_1(\lambda'(e \vert \lambda)]=\overline{v}_1(\lambda)$. 
Truth-telling is then induced if, given the current promise $\lambda$, the renewed promise $\lambda'$ satisfies the following incentive constraints.
\begin{align}
(1-\beta)u (e+\tau^*(e \vert \lambda)) &+ \beta \overline{v}_1(\lambda'(e \vert \lambda)) \nonumber \\
&\geq (1-\beta)u (e+\tau^*(\tilde{e} \vert \lambda)) + \beta \overline{v}_1(\lambda'(\tilde{e} \vert \lambda))  \text{ for }e, \tilde{e} \in E .\label{ineq:IC}
\end{align}

To construct a continuation promise satisfying both \eqref{ineq:PK} and
\eqref{ineq:IC}, 
define the deviation of the planner's continuation payoff from its average by
\[
\Delta v_2(\lambda,e):=v_2(\lambda,e)-\overline v_2(\lambda).
\]
Since $v_2$ is the discounted value of $-\tau$, a larger value of $\Delta v_2(\lambda,e)$
corresponds to a smaller current transfer (or a larger current contribution).
We choose $\lambda'(e\mid\lambda)$ so that this deviation is exactly offset by a change
in promised continuation utility:
\[
\beta\lambda\bigl(v_1(\lambda'(e\mid\lambda))-v_1(\lambda)\bigr)
=
\Delta v_2(\lambda,e).
\]
Equivalently,
\begin{align*}
\lambda\Bigl[(1-\beta)u(e+\tau^*(e\mid\lambda))&+\beta \overline{v}_1(\lambda'(e\mid\lambda))\Bigr] \\
&=
\lambda\Bigl[(1-\beta)u(e+\tau^*(e\mid\lambda))+\beta \overline{v}_1(\lambda)\Bigr] 
+v_2(\lambda,e)-\overline{v}_2(\lambda).
\end{align*}
Hence, the continuation payoff under truthful reporting coincides with the weighted-sum
problem used in \eqref{eq:PO}. Moreover, since
\[
E[\Delta v_2(\lambda,e)\mid \lambda]=0,
\]
taking expectations yields the promise-keeping condition.

The remaining issue is feasibility. Since $\overline{v}_1$ is bounded above by
\[
S:=\sup_{x\in\mathbb R_+}u(x),
\]
the inverse $\overline{v}_1^{-1}$ is well defined only if the target continuation utility remains
below $S$. If
\[
\overline{v}_1(\lambda)+\frac{1}{\beta\lambda}\bigl(v_2(\lambda,e^M)-\overline{v}_2(\lambda)\bigr)<S,
\]
where $e^M$ is the maximal income realization, then $\lambda'(e\mid\lambda)$ is
well defined for every $e\in E$, and we may define
\begin{align}
\lambda'(e\mid\lambda)
=
\overline{v}_1^{-1}\!\left(
\overline{v}_1(\lambda)+\frac{1}{\beta\lambda}\bigl(v_2(\lambda,e)-\overline{v}_2(\lambda)\bigr)
\right).\label{eq:lambda'}
\end{align}
 
On the other hand,  if there is an income $\hat{e} \in E$ for given $\lambda$ such that $\lambda'$ defined in \eqref{eq:lambda'} must promise the expected utility for the next period above the supremum of the individual's utility function, then the planner must find the closest promise for which there is a certainty equivalent to the promised level of lifetime utility. 
That is, for given $\lambda$, if there is  $\hat{e} \in E$ such that $\overline{v}_1(\lambda)+\lambda^{-1}\beta^{-1}(v_2(\lambda, \hat{e})-\overline{v}_2(\lambda)) \geq S$, 
for reported income $e \in E$, define $\lambda'$ such that  
\begin{align}
\lambda'(e\vert \lambda)=\overline{v}_1^{-1}\bigg(\overline{v}_1(\lambda) + \lambda^{-1}\beta^{-1}(v_2(\lambda, \min\{e, \overline{e}_{\lambda}\}) - \E_{\tau^{\ast}}\left[v_2(\lambda, \min\{e, \overline{e}_{\lambda}\})\right] \bigg), \label{eq:lambda'-near-e}
\end{align}
where $\E_{\tau^{\ast}}$ denotes
\[
\E_{\tau^{\ast}}\left[v_2(\lambda, \min\{e, \overline{e}_{\lambda}\})\right] := 
(1-\beta)\E\left[
-\tau^{\ast} (\min\{e_t,\overline{e}_{\lambda}\} \vert \lambda) +
\sum_{s=t+1}^{\infty}\beta^{s-t} (-\tau^{\ast} (e_s \vert \lambda_s))
\right]
\]
$\overline{e}_{\lambda}$ is defined such that 
\[
\overline{e}_{\lambda}=\arg\max_{e'\in E} 
\big(
v_2(\lambda,e')-
\E_{\tau^{\ast}}[v_2(\lambda,\min\{e_t,e'\})]
\big)
\]subject to 
\[
\overline{v}_1(\lambda)+\frac{1}{\lambda \beta}\big(
v_2(\lambda,e')-
\E_{\tau^{\ast}}[v_2(\lambda,\min\{e_t,e'\})]
\big)<S.
\]
This adjustment also changes the transfers in the current period $t$. However, we see the recursive relation in \eqref{eq:lambda'-near-e} satisfies the following corresponding incentive constraint. 
\begin{align}
(1-\beta)u (e&+\tau^*(\min\{e,\overline{e}_{\lambda}\} \vert \lambda)) + \beta \overline{v}_1(\lambda'(e \vert \lambda)) \nonumber \\
&\geq (1-\beta)u (e+\tau^*(\min\{\tilde{e},\overline{e}_{\lambda}\} \vert \lambda)) + \beta \overline{v}_1(\lambda'(\tilde{e} \vert \lambda))  \text{ for }e, \tilde{e} \in E .\label{ineq:IC'}
\end{align}
This is because the transfer $\tau^*(\min\{e,\overline{e}_{\lambda}\} \vert \lambda)$ in \eqref{ineq:IC'} is the solution to the problem \eqref{eq:PO} subject to $\tau(e_t) \geq (u')^{-1}(\lambda^{-1})-\overline{e}_{\lambda}$ for $t \in \mathbb{N} \cup \{0\}$, where $\lambda=\lambda_0$. In other words, since the transfer to the planner is constant for individuals whose income is above the threshold $\overline{e}_{\lambda}$, they are indifferent between reporting and misreporting their true income. 

Concerning the promise-keeping constraint, it holds because the expectation of the left-hand side of the inequality in \eqref{ineq:IC'} is greater than or equal to the promised level of lifetime utility $\overline{v}_1(\lambda)$:
\[\E\big[(1-\beta) u(e + \tau^*(\min(e,\overline{e}_{\lambda}) \vert \lambda))+\beta \overline{v}_1(\lambda')\big] \geq \overline{v}_1(\lambda).\]

\section{Risk-shifting mechanism}\label{sec:mechanism-M}
This section builds and analyzes the transfer rule $\mathcal{M}$. We first show that each individual's promised utility forms a bounded martingale, and then derive conditions under which the planner's inter-temporal budget is non-negative. 

Since the definition of $\lambda'$ in \eqref{eq:lambda'-near-e} is an extension of \eqref{eq:lambda'}, we can use it for the case where $\lambda'$ could also be defined in \eqref{eq:lambda'}. We define a mechanism like this.

A sufficient condition for the sequence of promise $\{\lambda_t\}_{t \in \mathbb{N}\cup \{0\}}$ to always satisfy the recursive condition \eqref{eq:lambda'} is, as we see in the following lemma, that the Arrow-Pratt measure of the absolute risk aversion of $u$ at the optimal consumption $(u')^{-1}(\lambda^{-1})$ is sufficiently low for all $\lambda>0$. 
\begin{assumption}\label{ass:moderate-risk}
(Moderate risk aversion). For all $\lambda>0$, 
\begin{align}
 -\frac{u''}{u'}((u')^{-1}(\lambda^{-1})) < \frac{\beta}{(1-\beta)(e^M-\E[e_t])}. \label{ineq:sufficient-condition-less-than-S}
\end{align}
\end{assumption}
\noindent
\emph{Interpretation.}
Assumption \ref{ass:moderate-risk} compares Arrow-Pratt absolute risk aversion at the
contract consumption level with a scaled income spread. A convenient class of examples is
given by utility functions with nonincreasing absolute risk aversion, since in that case the
condition is easiest to verify at low consumption levels. In Section \ref{sec:numerical_example} our numerical specification
\[
u(c)=\frac{(c+1)^{1+\gamma}}{1+\gamma}, \qquad \gamma=-3,
\]
has
\[
-\frac{u''(c)}{u'(c)}=\frac{3}{c+1},
\]
so absolute risk aversion is decreasing.

We define risk-shifting mechanism as follows. 
\begin{definition}
(Risk-shifting mechanism)
A risk-shifting mechanism $\mathcal{M}$ is 
 a sequence of promises $\{\lambda_t\}_{t \in \mathbb{N}\cup \{0\}}$ and transfers 
$\tau_t(e_t)=\tau^*(\min\{e_t, \overline{e}_{\lambda_t}\} \vert \lambda_{t})$ such that $\lambda_{t+1}$ 
is given by \eqref{eq:lambda'-near-e}.  
\end{definition}

\begin{remark}
Under Assumption \ref{ass:moderate-risk}, the feasibility correction
never binds, so current consumption is constant at
\[
c_t=(u')^{-1}(\lambda_t^{-1}),
\]
and the realized shock affects only the next-period promise. If the feasibility correction
\eqref{eq:lambda'-near-e} binds, then agents whose unconstrained continuation promise
would exceed the maximal feasible utility level receive the capped continuation promise
instead. Since their contribution is capped at the threshold level, they retain part of the
current income shock in current consumption. Hence, full within-period insurance fails for
those realizations, but the promise-keeping condition holds weakly in discounted lifetime
utility.
\end{remark}

\begin{remark}
If utility were unbounded above, the feasibility issue behind Assumption \ref{ass:moderate-risk} would disappear, but the martingale-convergence argument would then require a different integrability condition.
\end{remark}

 \begin{lemma}\label{lem:sufficient-condition-less-than-S}
Under Assumption \ref{ass:moderate-risk} 
 the sequence of promises $\{\lambda_t\}_{t \in \mathbb{N}\cup \{0\}}$ is always defined in \eqref{eq:lambda'}. 
 \end{lemma}
\noindent
\emph{Proof: see Appendix \ref{sec:app-prop1-2}}.

If the inequality \eqref{ineq:sufficient-condition-less-than-S} does not hold, there may be a case where a promised-utility state $\lambda_t$ in $\mathcal{M}$ is defined in \eqref{eq:lambda'-near-e} instead of in \eqref{eq:lambda'}, e.g. if the individuals' utilities have constant absolute risk aversion, and if it is greater than $\beta/((1-\beta)(e^M-\E[e_t]))$.  

The risk-shifting mechanism $\mathcal{M}$ is sequentially incentive compatible regardless of whether each $\lambda_t$ is defined in \eqref{eq:lambda'} or in \eqref{eq:lambda'-near-e}. Furthermore, if $\{\lambda_t\}_{t \in \mathbb{N}\cup \{0\}}$ is always defined in \eqref{eq:lambda'}, like the $\lambda$ mechanism of \cite{Marcet-Marimon-1992}, $\mathcal{M}$ is sequentially efficient.  
\begin{theorem}\label{thm:sequential_IC}
The mechanism $\mathcal{M}$ defined by $(\tau^{\ast}, \{\lambda_t\})$ is 
 sequentially incentive compatible. Under Assumption \ref{ass:moderate-risk}, it is 
 sequentially efficient.% and the promised utility process $\{\overline{v}_1(\lambda_t)\}$ is a bounded martinga
\end{theorem}
\noindent
\emph{Proof: see Appendix \ref{sec:app-prop1-2}}.

\subsection{Individual level: bounded promised utilities}\label{sec:long-term-properties}
This subsection shows that the sequence of promised utilities forms a martingale and therefore converges, so immiseration cannot occur.

Let $(\mathcal{F}_t)$ be the natural filtration generated by $\{\lambda_s\}_{s \leq t}$. Because \eqref{eq:lambda'-near-e} makes $\lambda_{t+1}$ measurable w.r.t. the $\sigma$-algebra generated by $\lambda_t$ and $e_t$, the process $\overline{v}_1(\lambda_t)$ is adapted. 
\begin{theorem}\label{thm:main}
The promised utility process {\small$\{\overline{v}_1 (\lambda_{n})\}_{n \in \mathbb{N}\cup \{0\}}$} induced by  $\mathcal{M}$ is a martingale and 
converges to an integrable random variable a.s. (no immiseration).  
\end{theorem}
\noindent
\emph{Proof: see Appendix \ref{sec:app-prop1-2}}.

We have for $n\in \mathbb{N}$ $\E[\overline{v}_1(\lambda_n) \vert \mathcal{F}_{n-1}] = \overline{v}_1(\lambda_{n-1})$ a.e. $[P]$. Hence, $\{\overline{v}_1(\lambda_{n-1})\}_{n \in \mathbb{N}}$ is a martingale, and 
promised lifetime utility is preserved in conditional expectation over time. Realized promised utility may vary across histories, but it exhibits no systematic drift.
Since the process is bounded, it converges almost surely to an integrable random variable,
so immiseration does not occur. In \cite{thomas-warrall-1990}, the state variable (promised utility) $V^t$ converges to $-\infty$, whereas the martingale object is the lender's marginal value process $U'_*(V^t)$. 

\subsection{Planner's budget and sustainability}\label{sec:sustainability}
All claims in this subsection are sufficient (not necessary) conditions based on a second-order expansion.
We derive sufficient conditions under which the cross-sectional budget is non-negative each period. 

The convergence of individuals' lifetime utilities to an integrable random variable with finite expectation does not yet ensure the sustainability of the  $\mathcal{M}$ mechanism. This is because the planner predicts a declining trend in its budget for each individual in the long run, as we see in the following lemma. 
Let $\overline{\tau}_n$ denote the average transfer to an $n$-period-old individual (positive means the planner pays). Then $\overline{\tau}_n$ is:
\begin{align*}
\overline{\tau}_n &=\E[\tau^{\ast}(e_n \vert \lambda_n)]=\E [u^{-1}(\overline{v}_1(\lambda_n))] - \E[e_n] \quad \text{ for }n \in \mathbb{N}\cup\{0\}. 
\end{align*}
\begin{lemma}\label{lem:Jensen}
The average transfer for each cohort $\{\overline{\tau}_n\}_{n \in \mathbb{N} \cup \{0\}}$ is an increasing sequence.
\end{lemma}
\noindent
\emph{Proof sketch} (details in Appendix \ref{sec:app-jen}).
For $n=1$, since $\E[\overline{v}_1(\lambda_1(e_0 \vert \lambda_0))]=\overline{v}_1(\lambda_0)$ and $u^{-1}$ is convex from Jensen's inequality we have 
\begin{align*}
\overline{\tau}_1 &= \E[u^{-1}(\overline{v}_1(\lambda_1(e_0 \vert \lambda_0)))] - \E[e_1] \\
& \geq u^{-1}(\overline{v}_1(\lambda_0)) - \E[e_0] = \overline{\tau}_0.
\end{align*}
The proof of Lemma \ref{lem:Jensen} shows that the increasing trend in $\{\overline{\tau}_n\}_{n \in \mathbb{N}\cup \{0\}}$ is due to the cost of keeping the expected lifetime utility in each period unchanged from the previous period, although there are risks between periods.

Let $g_t(\lambda_0)$ be the expected budget balance in period $t$ for the planner for a given initial promise $\lambda_0$. 
Since each individual's probability of being alive is $\alpha \in (0,1)$, and from period one the next generation of continuum of individuals on the $(1-\alpha)$ interval is born in each period, 
$g_t(\lambda_0)$ is given by 
\begin{align*}
g_t(\lambda_0)&=\sum_{k=1}^t \alpha^k (\overline{\tau}_{k-1}-\overline{\tau}_k) -\overline{\tau}_0 .
\end{align*}
Here, a positive $g_t$ denotes a surplus (the planner collects funds). 
For the $\mathcal{M}$ mechanism to be sustainable, the budget balance in each period must be greater than or equal to zero. As we see in the following proposition, under a second-order (Taylor) approximation, if incomes have a symmetric distribution, 
a sufficient condition for the budget balance in period $t$ to be greater than or equal to be zero is that (i) the absolute risk aversion of $u$ is decreasing in consumption, (ii) the absolute risk aversion of the absolute risk aversion of $u$ is less than the absolute risk aversion of $u^{-1}$ and (iii) $\lambda_0$ is set so that the inequality $(\overline{\tau}_1-\overline{\tau}_0)\alpha/(1-\alpha) \leq -\overline{\tau}_0$ is satisfied.

\begin{proposition}[second-order sufficient condition for budget balance]\label{prop:sustainablity}
In a stationary overlapping-generations society with i.i.d. income risk, if 
\begin{itemize}
\setlength\itemsep{2pt}
\item[\normalfont (i)] incomes have a symmetric distribution, 
\item[\normalfont (ii)] $-u''/u'$ is a decreasing function, 
\item[\normalfont (iii)] $(-u''/u') \circ u^{-1}$ is a concave function
and 
\item[\normalfont (iv)] initial condition $\alpha(\overline{\tau}_1 -\overline{\tau}_0)/(1-\alpha) \leq -\overline{\tau}_0$ is satisfied,  
\end{itemize}
then planner's period-by-period balance 
$g_t(\lambda_0) \geq 0$ for $t \in \mathbb{N}$ under a second-order (Taylor) approximation. 
\end{proposition}

\noindent
\emph{Proof sketch.}
The key step is to show that the second-order approximation of the average-transfer
increment $\overline\tau_n-\overline\tau_{n-1}$ is decreasing in $n$.
For a fixed current promise $\lambda_{n-1}$ and realization $e$, define
\[
\Delta_{n-1}
:= \overline v_1(\lambda_n(e\mid\lambda_{n-1}))-\overline v_1(\lambda_{n-1})
= \frac{1-\beta}{\beta\lambda_{n-1}}(e-E[e_n]).
\]
We expand
\[
u^{-1}(\overline v_1(\lambda_n(e\mid\lambda_{n-1})))
-u^{-1}(\overline v_1(\lambda_{n-1}))
\]
around $\overline v_1(\lambda_{n-1})$. As shown in Appendix~\ref{sec:app-sus}, a
third-order Taylor expansion yields a second-order approximation for
$\overline\tau_n-\overline\tau_{n-1}$ after taking expectations. In the symmetric case,
the odd-order term cancels in expectation, so the leading term is the second-order term.
Assumptions (ii) and (iii), together with Jensen's inequality, imply that this second-order
term is decreasing in $n$. Hence the second-order approximation of
$\{\overline\tau_n-\overline\tau_{n-1}\}_{n\ge1}$ is decreasing.

Therefore,
\[
\overline\tau_n-\overline\tau_{n-1}\le \overline\tau_1-\overline\tau_0
\qquad\text{(under the second-order approximation),}
\]
and substituting this bound into
\[
g_t(\lambda_0)=\sum_{k=1}^t \alpha^k(\overline\tau_{k-1}-\overline\tau_k)-\overline\tau_0
\]
shows that condition (iv) is sufficient for $g_t(\lambda_0)\ge0$ for all $t\in\mathbb N$.
See Appendix~\ref{sec:app-sus} for the complete proof.

\section{Numerical examples}\label{sec:numerical_example}
An intuitive sufficient condition for $\lambda_0$ to satisfy the initial condition $\alpha(\overline{\tau}_1 -\overline{\tau}_0)/(1-\alpha) \leq -\overline{\tau}_0$ specified in Proposition \ref{prop:sustainablity} 
 is given by the second-order polynomial approximation as follows. 
\begin{corollary}\label{cor:sufficient}
The condition $\alpha(\overline{\tau}_1 -\overline{\tau}_0)/(1-\alpha) \leq -\overline{\tau}_0$ in Proposition \ref{prop:sustainablity} holds with  the second-order polynomial approximation if the initial promise $\lambda_0$ satisfies
\begin{align}
\lambda_0 (\overline{v}_1(\lambda_E)-\overline{v}_1(\lambda_0)) \geq 
\sqrt{
\frac{(1-\alpha r)^2}{\alpha r^2 (1-\alpha)}\Var[e_t]},\label{ineq:sufficient-lambda_0}
\end{align}
where $\lambda_E=1/(u'(\E[e_t]))$ denotes the promise corresponding to mean income report. 
\end{corollary}
\noindent
\emph{Proof: see Appendix \ref{sec:appen-cor}}. 
\begin{figure}[H]
\begin{center}
\includegraphics[width=8cm]{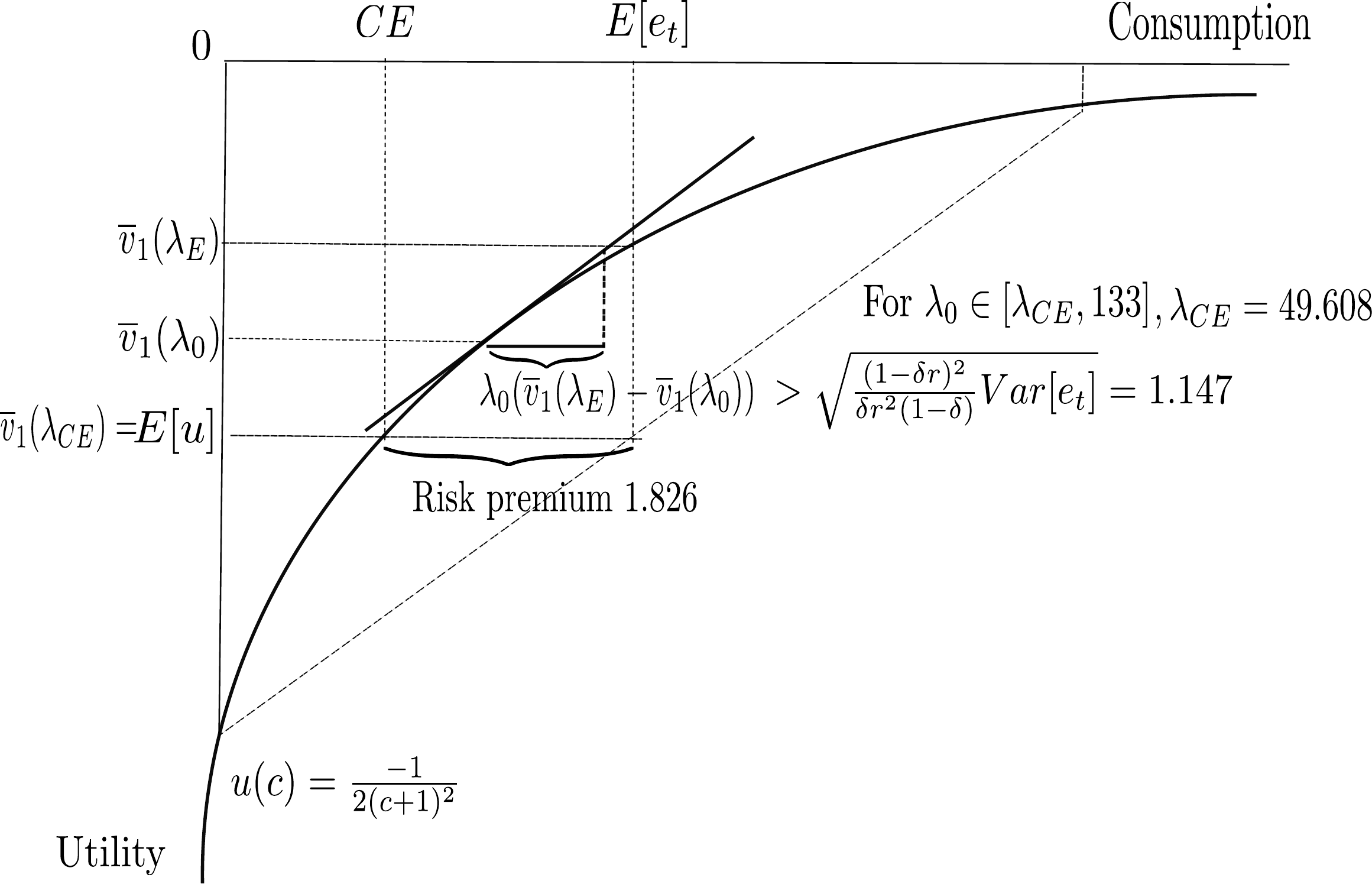} %\\ \vspace{3pt}
\caption{Utility Function}
\label{fig:utility_function}
\end{center}
\end{figure}\vspace{-12pt}
{\footnotesize 
\begin{singlespace*}
\noindent
Notes: This figure illustrates a utility function (vertical axis) against consumption (horizontal axis). The segment $\lambda_0(\overline{v}_1(\lambda_E)-\overline{v}_1(\lambda_0))$ appears in Corollary \ref{cor:sufficient}. 
It is the linear approximation of the length between the expected income and the consumption guaranteed in the mechanism M with $\lambda_0$. 
\end{singlespace*}
}
\vspace{12pt}

The left-hand side of the inequality \eqref{ineq:sufficient-lambda_0} is a linear approximation of the difference between the expected income and the consumption level in the mechanism $\mathcal{M}$ with the promise $\lambda_0$.  The inequality \eqref{ineq:sufficient-lambda_0} means that this length is greater than the square root of the weighted income variance. Figure \ref{fig:utility_function} shows which part of the graph of a utility function corresponds to  $\lambda_0(\overline{v}_1(\lambda_E)-\overline{v}_1(\lambda_0))$.
The lifetime utility $\overline{v}_1(\lambda_0)$ in the mechanism $\mathcal{M}$ is higher than that in the state of autarky if the risk premium of the state of autarky is higher than the term in the right-hand side of the inequality \eqref{ineq:sufficient-lambda_0}. 

A higher discount factor and higher probability of being alive will relax the sufficient condition of $\lambda_0$ for making the mechanism $\mathcal{M}$ sustainable. If $(u^{-1})'''$ is positive, the condition \eqref{ineq:sufficient-lambda_0} may be unnecessarily strict. 

We illustrate the mechanism with a Laurent polynomial utility, a symmetric discrete income distribution, and the benchmark parameter $r=\alpha=0.93$. 
\begin{itemize}
\setlength\itemsep{2pt}
\item Utility $u(c)=\frac{1}{1+\gamma}(c+1)^{1+\gamma}$, $\gamma=-3$
\item Income support $E=\{0,1,2,3,4,5,6,7,8,9\}:=E_s$
\item Probabilities $P(e)$ (mirror-symmetric, approx. normal) $:=P_s$
\item Discount factor and survival rate $r=\alpha=0.93$
\item Initial state $\lambda_0=50 \in  [\lambda_{CE},133]$
\end{itemize}
The values follow Corollary \ref{cor:sufficient}: in this calibration the lower bound $\lambda_{CE}=1/u'\bigl(u^{-1}(E[u(e_t)])\bigr)=49.608$ ensures higher lifetime utility than autarky across all cohorts, and the upper bound $133$ is set to satisfy \eqref{ineq:sufficient-lambda_0}. 

\begin{figure}[H]
\begin{center}
\includegraphics[width=8cm]{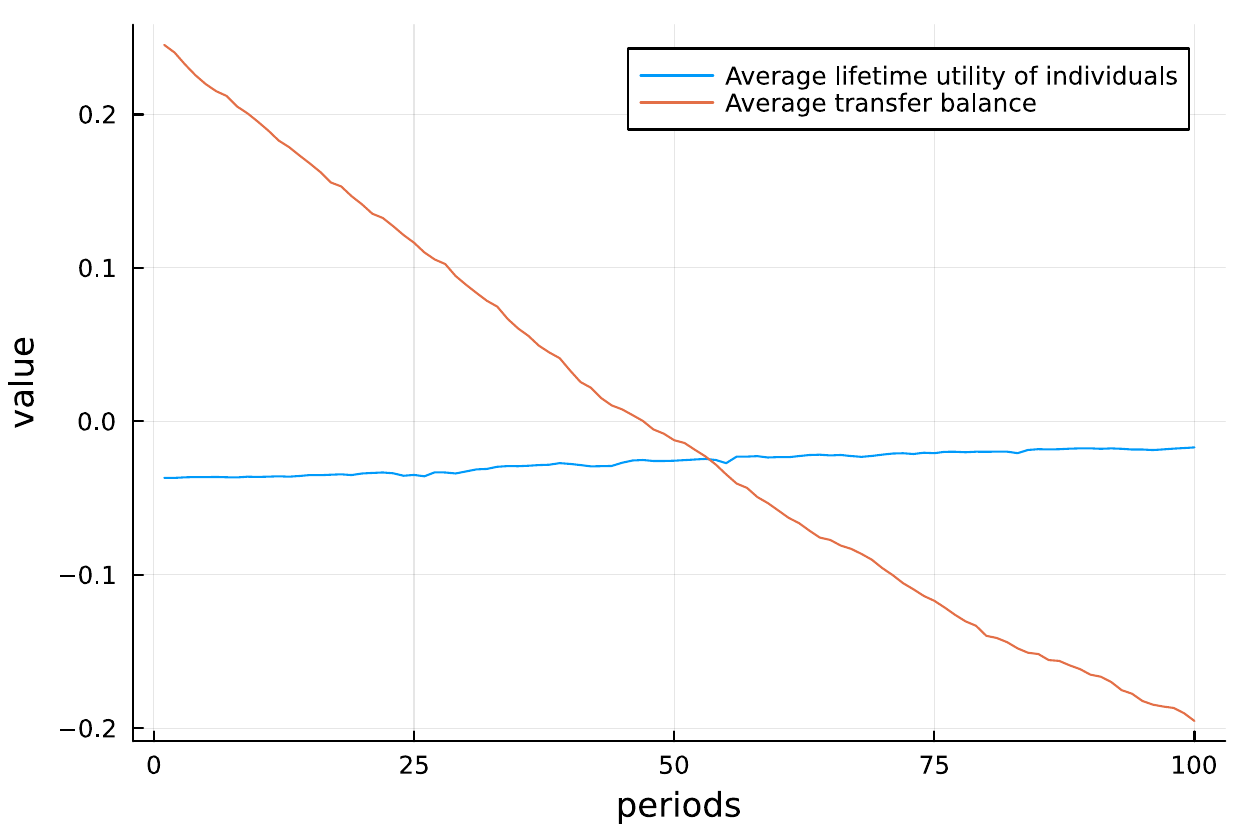}
\caption{Sample mean of lifetime utilities and transfer balances}
\label{fig:plot_average_lifetime_utiity_transfer}
\end{center}
\end{figure}

Figure \ref{fig:plot_average_lifetime_utiity_transfer} shows a series of sample means of lifetime utilities of $1000$ individuals of the same generation and corresponding transfer balances for $100$ periods for the above case. 
The initial promise is set  $\lambda_0=50$. 
Consistent with Theorem \ref{thm:main}, the utility path stays bounded while transfers 
$\{-\overline{\tau}_n\}_{n=0}^{100}$ fall, echoing Lemma \ref{lem:Jensen}. 

\begin{figure}[H]
\begin{center}
\includegraphics[width=8cm]{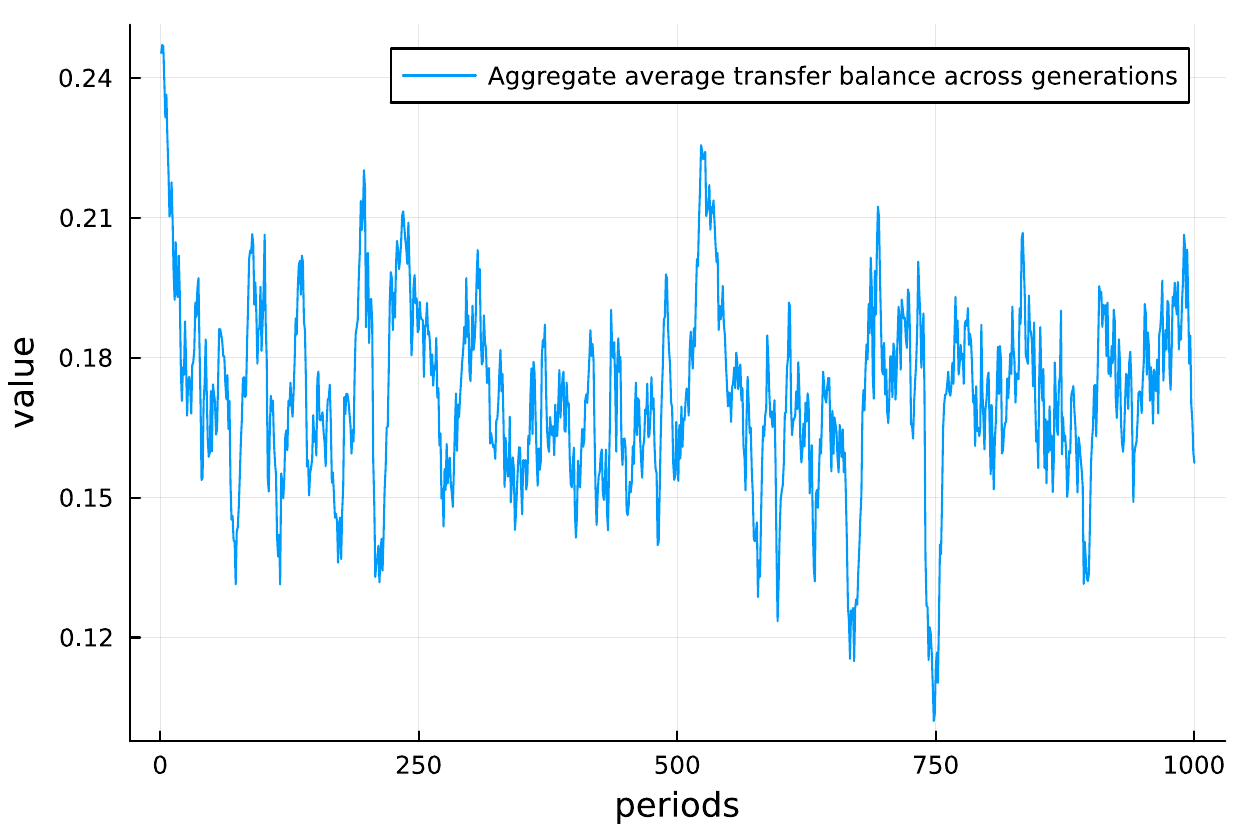} %\\ \vspace{3pt}
\caption{Sample means of budget balance $\{g_t(\lambda_0)\}$}
\label{fig:plot_aggregate_average_transfer_balance}
\end{center}
\end{figure}

Figure \ref{fig:plot_aggregate_average_transfer_balance} shows a series of sample means of transfer balances for a society of $100$ individuals for $1000$ periods. In this society, each individual will be alive in the next period with probability $\alpha=0.93$, and newly born individuals enter the society, holding the total population $100$. As in Figure \ref{fig:plot_average_lifetime_utiity_transfer}, each individual's utility is $u(c)=(c+1)^{1+\gamma}/(1+\gamma)$ with $\gamma=-3$, and  the parameters are the initial promise $\lambda_0=50$ and the discount factor $r=0.93$. The income set $E$ is $E_s$. The random numbers are generated with the same symmetric distribution $P_s$.

The series of sample means of transfer balances $\{g_t(\lambda_0)\}_{t=0}^{1000}$ remains non-negative, and is consistent with  
 Proposition \ref{prop:sustainablity}. 

\section{Dynamic-programming verification}\label{sec:Bellman}
This section verifies that the transfer rule $\tau^{\ast}$ 
 solves the infinite-horizon Bellman equation and that $V$ is the corresponding value function (under condition \eqref{ineq:sufficient-condition-less-than-S} of Lemma \ref{lem:sufficient-condition-less-than-S}).

For a given current promise $\lambda$, let $\Lambda(\lambda)$ denote the set of
measurable controls $\tau:E\to\mathbb R$ such that the implied continuation
promise $\lambda'_\tau(e\mid\lambda)$
defined by \eqref{eq:lambda'} whenever feasible, and by \eqref{eq:lambda'-near-e}
otherwise, is well defined for all $e\in E$, and the pair $(\tau,\lambda'_\tau)$ 
satisfies the promise-keeping condition \eqref{ineq:PK} and the sequential incentive
constraint \eqref{ineq:IC}.

The Bellman equation (DP) is given by
\begin{align}
V(\lambda_t,t) = \sup_{\tau \in \Lambda(\lambda_t)} \{(1-\beta)u(e_t + \tau(e_t \vert \lambda_t))+ \beta \E_t[V(\lambda'_\tau(e_t\vert \lambda_t), t+1)]\}. \label{eq:Bellman}
\end{align}
Under Assumption \ref{ass:moderate-risk}, $\lambda'_\tau$ is given by \eqref{eq:lambda'}. 

Our candidate solution is $V$ in \eqref{eq:V} with control $\tau^{\ast}$.\footnote{Although we write $V(\lambda_t,t)$ to match the terminal-condition formulation of
\cite{wiszniewska2011terminal}, the economic environment is stationary; the time
argument is used only for the dynamic-programming verification. Along the candidate
mechanism, $V(\lambda_t,t)=v_1(\lambda_t)$.} 
\begin{align}
V(\lambda_t, t)=\E_t \big[ u(e_t + \tau^*(e_t \vert \lambda_t))\big], \quad \text{for }\, \lambda_t \in \{\lambda_t\}_{t \in \mathbb{N}\cup \{0\}}. \label{eq:V}
\end{align}

We observe the following proposition.  
\begin{proposition}\label{prop:bellman}
With Assumption \ref{ass:moderate-risk}, $V$ in \eqref{eq:V} solves the infinite-horizon Bellman equation, and $\tau^*$ maximizes the lifetime value $J$ in 
\eqref{eq:def_of_J}. 
\end{proposition}

\begin{proof}[Proof sketch]
Step 1: we verify $V$ in \eqref{eq:V} satisfies DP by substituting $\tau^{\ast}$. Step 2: we check a sufficient condition for $V$ to be the value function of the Bellman equation, which is given by  \cite{wiszniewska2011terminal}.\footnote{ {\bf Terminal condition} \citep[][]{wiszniewska2011terminal} ~
 For every $\lambda_t \in \{\lambda_t\}_{t \in \mathbb{N} \cup \{0\}}$ in $\mathcal{M}$ \   
$\lim_{t \to \infty} V(\lambda_t,t) \beta^t \leq 0$
and 
 for every $\lambda_t \in \{\lambda\}_{t \in \mathbb{N} \cup \{0\}}$ in $\mathcal{M}$, if $\lim_{t \to \infty} V(\lambda_t,t) \beta^t <0$, then 
$J(\lambda_t,t, \hat{\tau}) =-\infty $
for every $\hat{\tau}: \mathbb{R}_{++} \to \mathbb{R}_+$ such that $\lambda_t$ in $\mathcal{M}$ corresponds to $\hat{\tau}$.} 
Step 3: we conclude that $\tau^{\ast}$ maximizes $J$ for any initial period $t$ and for $\lambda_t \in \{\lambda_k\}_{k=t}^{\infty}$.
\end{proof} (complete proof in Appendix \ref{sec:app-bellman}).

\section{Conclusion}\label{sec:conclusion}
This paper shows that immiseration is not unavoidable within the risk-shifting mechanism studied here. The mechanism induces a bounded promised-utility process and, under the paper's sufficient condition, is sequentially efficient. In the stationary OLG extension, the paper also identifies sufficient conditions under which intergenerational transfers sustain the mechanism and can improve on autarky. 

\begin{appendices}
\section{Proof of theorems 1 and 2}\label{sec:app-prop1-2}
\counterwithin{theorem}{section}
\counterwithin{lemma}{section}
\counterwithin{proposition}{section}
\numberwithin{equation}{section}
\begin{proof}[Proof of Lemma \ref{lem:sufficient-condition-less-than-S}]
Write $F(\lambda)=\overline{v}_1(\lambda)+\lambda^{-1}\beta^{-1}(v_2(\lambda, e^M)-\overline{v}_2(\lambda))$ for $\lambda>0$.
It suffices to show the proposition that $F(\lambda) <S$ for all $\lambda>0$. 

  Since the utility function $u$ is continuous, bounded from above and satisfies the Inada condition, we have 
 \begin{align*}
 &\lim_{\lambda \to \infty} \overline{v}_1(\lambda)=u(\lim_{\lambda\to \infty} (u')^{-1}(\lambda^{-1})) \leq S \\
&\lim_{\lambda \to \infty} (u')^{-1}(\lambda^{-1}) = \infty.
  \end{align*}
 So we see that 
 \[\lim_{\lambda \to \infty} F(\lambda)=\lim_{\lambda\to \infty} \overline{v}_1(\lambda)=S.\]
 
 If \eqref{ineq:sufficient-condition-less-than-S} holds, then we have $F'(\lambda) > 0$ for $\lambda>0$, which implies that $F$ is strictly increasing in $\lambda$.  
 Consequently, we see the desired result $F(\lambda) <S$ for all $\lambda>0$.
 \end{proof}

\vspace{12pt}
\begin{proof}[Proof of Theorem \ref{thm:sequential_IC}]
Similar proof can be found in \cite{Marcet-Marimon-1992}. 
%For completeness, we report the entire proof. 

We first show that $\mathcal{M}$ is sequentially incentive compatible. For 
$\lambda_t>0$ with $F(\lambda_t) <S$
\begin{align*}
\lambda_t \big[&(1-\beta)u(e+\tau^*(e \vert \lambda_t))+\beta \overline{v}_1(\lambda_{t+1}(e\vert \lambda_t))\big] \\
&\underset{\eqref{eq:lambda'}}{=}\lambda_t \big[(1-\beta)u(e+\tau^*(e \vert \lambda_t))+\beta \overline{v}_1(\lambda_t)\big]+v_2(\lambda_t,e)- \overline{v}_2(\lambda_t) \\
&\geq \lambda_t \big[(1-\beta)u(e + \tau^*(\tilde{e}\vert \lambda_t)) + \beta \overline{v}_1(\lambda_t) \big] + v_2(\lambda_t, \tilde{e}) - \overline{v}_2(\lambda_t) \\
&=\lambda_t \big[(1-\beta)u(e+\tau^*(\tilde{e}\vert\lambda_t))+\beta \overline{v}_1(\lambda_{t+1}(\tilde{e}\vert \lambda_t))\big] \text{ for }e, \tilde{e} \in E.
\end{align*}

The last inequality follows from the optimality of $\lambda \overline{v}_1(\lambda) + v_2(\lambda, e)$ in the problem \eqref{eq:PO} given $\lambda$.
Hence, $\mathcal{M}$ is sequentially incentive compatible for  $\lambda>0$ defined in \eqref{eq:lambda'}.

For $\lambda_t>0$ with $F(\lambda_t) \geq S$ 
\begin{align*}
\lambda_t \big[(1-\beta)&u(e+\tau^*(\min\{e,\overline{e}_{\lambda_t}\} \vert \lambda_t))+\beta \overline{v}_1(\lambda_t'(e\vert \lambda_t))\big] \\
&\underset{\eqref{eq:lambda'-near-e}}{=}\lambda_t \big[(1-\beta)u(e+\tau^*(\min\{e,\overline{e}_{\lambda_t}\} \vert \lambda_t))+\beta \overline{v}_1(\lambda_t)\big] \\ & \hspace{86pt} + v_2(\lambda_t,\min\{e,\overline{e}_{\lambda_t}\})- \E_{\tau^*}[v_2(\lambda_t,\min\{e_t,\overline{e}_{\lambda_t}\})] \\
&\geq \lambda_t \big[(1-\beta)u(e + \tau^*(\min\{\tilde{e},\overline{e}_{\lambda_t}\}\vert \lambda_t))
+ \beta \overline{v}_1(\lambda_t) \big] \\ &  
\hspace{86pt} +v_2(\lambda_t, \min\{\tilde{e},\overline{e}_{\lambda_t}\}) - \E_{\tau^*}[v_2(\lambda_t, \min\{e_t,\overline{e}_{\lambda_t}\})] \\
&=\lambda_t \big[(1-\beta)u(e+\tau^*(\min\{\tilde{e},\overline{e}_{\lambda_t}\}\vert\lambda_t))+\beta \overline{v}_1(\lambda_t'(\tilde{e}\vert \lambda_t))\big] \text{ for }e, \tilde{e} \in E.
\end{align*}
The last inequality follows from the fact that $\tau^*(\min\{e,\overline{e}_{\lambda_t}\}\vert \lambda_t)$ corresponds to the solution of the problem \eqref{eq:PO} (planner's static problem at given $\lambda$) subject to $\tau_t \geq (u')^{-1}(\lambda^{-1}_0)-\overline{e}_{\lambda_0}$ for $t \in \mathbb{N}$, where $\lambda_0=\lambda_t$. 

\vspace{12pt}

We next show that, under condition
\eqref{ineq:sufficient-condition-less-than-S}, the continuation rule induced by
$\mathcal{M}$ solves problem \eqref{eq:PO} at every feasible current state. 
By the first part of this proof, $\mathcal{M}$  is sequentially incentive compatible.
 By Lemma \ref{lem:sufficient-condition-less-than-S} since condition \eqref{ineq:sufficient-condition-less-than-S} holds,  the sequence of promises $\{\lambda_t\}_{t \in \mathbb{N}\cup \{0\}}$ in  $\mathcal{M}$ always defined by \eqref{eq:lambda'}. 
 Hence $\tau^*$ in $\mathcal{M}$ corresponds to transfers that solve the problem \eqref{eq:PO} in every period. 
Therefore, conditional on the current promised state and within the admissible class where \eqref{eq:lambda'} applies, $\tau^*$ solves the planner's static problem at each date and induces a Pareto-efficient continuation payoff pair.

It remains to prove that under condition \eqref{ineq:sufficient-condition-less-than-S} $\mathcal{M}$ is not Pareto dominated by any other sequentially incentive compatible mechanism.  Suppose, contrary to our claim, that under Assumption \ref{ass:moderate-risk} there exists a sequentially incentive compatible mechanism $\Gamma$ that Pareto dominates $\mathcal{M}$ for a given income state $e$. Let $(v_1^*,v_2^*)$ denote the continuation payoff pair induced by $\Gamma$.
Since $v_1^*$ lies in the range of $\overline v_1$, define
\[
\lambda_0=\overline v_1^{-1}(v_1^*).
\]
Because condition \eqref{ineq:sufficient-condition-less-than-S} depends only on
the primitives and holds for every $\lambda>0$, it also holds at $\lambda_0$.
Hence $\mathcal M$ is well defined from the initial state $(\lambda_0,e)$, and
its continuation rule is given by \eqref{eq:lambda'}.
By construction, $\Gamma$ and $\mathcal M$ deliver the same continuation utility
to the agent, so $v_1^*=\overline v_1(\lambda_0)$. Since $\Gamma$ Pareto dominates
$\mathcal M$, it must satisfy
\[
v_2^*>v_2(\lambda_0,e).
\]
Hence
\[
\lambda_0 v_1^*+v_2^*>
\lambda_0 \overline v_1(\lambda_0)+v_2(\lambda_0,e),
\]
which contradicts the optimality of $\tau^*$ in problem \eqref{eq:PO} at
$(\lambda_0,e)$.
\end{proof}

\vspace{12pt}

\begin{proof}[Proof of Theorem \ref{thm:main}]
We first show that $\{\overline{v}_1(\lambda_{n})\}_{n \in \mathbb{N}\cup \{0\}}$ is a martingale. 
Since the sequence $\{\lambda_t\}_{t \in \mathbb{N} \cup \{0\}} \in \mathcal{M}$ satisfies \eqref{eq:lambda'-near-e}, we have  
\[\E[\overline{v}_1(\lambda_n)\vert \overline{v}_1(\lambda_{n-1})(\omega)]=\overline{v}_1(\lambda_{n-1})(\omega)\] for given realized $\overline{v}_1(\lambda_{n-1})(\omega)$. We thus have the following equalities: For all $B \in \mathcal{B}(\mathbb{R})$ for all $n \in \mathbb{N}$  
\begin{align*}\int_B\E\big[\overline{v}_1(\lambda_n)   \big\vert   \overline{v}_1(\lambda_{n-1})(\omega)\big] dP_{\overline{v}_1(\lambda_{n-1})}&=\int_B \overline{v}_1(\lambda_{n-1})(\omega) dP_{\overline{v}_1(\lambda_{n-1})}, 
\end{align*}
and 
\begin{align*}\int_B\E\big[\overline{v}_1(\lambda_n)   \big\vert   \overline{v}_1(\lambda_{n-1})(\omega)\big] dP_{\overline{v}_1(\lambda_{n-1})}&=\int_{\{\overline{v}_1(\lambda_{n-1})(\omega) \in B\}} \overline{v}_1(\lambda_n) dP \\
&= \int_{\{\overline{v}_1(\lambda_{n-1})(\omega) \in B\}} \overline{v}_1(\lambda_{n-1}) dP. 
\end{align*}
From the last equality we have for $n\in \mathbb{N}$ $\E[\overline{v}_1(\lambda_n) \vert \mathcal{F}_{n-1}] = \overline{v}_1(\lambda_{n-1})$ a.e. $[P]$, and $\{\overline{v}_1(\lambda_{n})\}_{n \in \mathbb{N}\cup \{0\}}$ is a martingale as claimed. 

We proceed to show that $\sup_{n}\overline{v}_1(\lambda_n)^+ < \infty$. From \eqref{eq:lambda'} and \eqref{eq:lambda'-near-e} we have 
$\overline{v}_1(\lambda_n) \leq S$ for $n \in \mathbb{N}$. So we have  $\sup_{n}\overline{v}_1(\lambda_n)^+ < \infty$.

Since $\{\overline{v}_1(\lambda_{n-1})\}_{n \in \mathbb{N}}$ is a martingale and $\sup_{n}\E[\overline{v}_1(\lambda_n)^+]\leq \sup_n \overline{v}_1(\lambda_n)^+<\infty$, 
from the submartingale convergence theorem, there is an integrable random variable $v^{\infty}$ such that $\overline{v}_1(\lambda_n) \to v^{\infty}$ almost everywhere.
\end{proof}

\section{Proof of Lemma \ref{lem:Jensen}}\label{sec:app-jen}
\begin{proof}[Proof of Lemma \ref{lem:Jensen}]
We first consider the case $n=1$. 
For $n=1$, since $\E[\overline{v}_1(\lambda_1(e_0 \vert \lambda_0))]=\overline{v}_1(\lambda_0)$ and $u^{-1}$ is convex from Jensen's inequality we have 
\begin{align*}
\overline{\tau}_1 &= \E[u^{-1}(\overline{v}_1(\lambda_1(e_0 \vert \lambda_0)))] - \E[e_1] \\
& \geq u^{-1}(\overline{v}_1(\lambda_0)) - \E[e_0] = \overline{\tau}_0.
\end{align*}
For $n>1$, since $u^{-1}$ is a convex function, from Jensen's inequality we see the result. 
\begin{align*}
\overline{\tau}_{n-1}&=\E[u^{-1}(\overline{v}_1(\lambda_{n-1}))] - \E[e_{n-1}]\\
&=\E[u^{-1}(\E[\overline{v}_1(\lambda_n) \vert \overline{v}_1(\lambda_{n-1})])]-\E[e_{n-1}]\\
&\leq \E[\E[ u^{-1}(\overline{v}_1(\lambda_n)) \vert \overline{v}_1(\lambda_{n-1})]] -\E[e_{n-1}] \\
&=\E[ u^{-1}(\overline{v}_1(\lambda_n))] -\E[e_{n}]\\
&=\overline{\tau}_n \quad \text{ for }n=2,3,\cdots.
\end{align*}
\end{proof}

\vspace{12pt}

\section{Proof of Proposition \ref{prop:sustainablity}}\label{sec:app-sus}
The proof of Proposition \ref{prop:sustainablity} uses the following lemma.
\begin{lemma}\label{lem:decreasing-sequence}
Under the assumptions (i), (ii), (iii) in Proposition \ref{prop:sustainablity}, 
the second-order polynomial approximation of 
$\overline{\tau}_n-\overline{\tau}_{n-1}$ is a decreasing sequence.
\end{lemma}
\begin{proof}
First we consider the case where $\{\lambda_n\}_{n \in \mathbb{N}\cup\{0\}}$ can be defined in  \eqref{eq:lambda'}. 
We use a third-order Taylor expansion to obtain the second-order approximation used in the proof of Proposition \ref{prop:sustainablity}. 
Using the third-order Taylor polynomial, for a given promise $\lambda_{n-1}>0$ and a given income $e \in E$, the corresponding transfer is written as
{\footnotesize 
\begin{align}
u^{-1}(\overline{v}_1(\lambda_n(e \vert \lambda_{n-1})))-u^{-1}(\overline{v}_1(\lambda_{n-1}))  =&
 \frac{1}{u'(u^{-1}(\overline{v}_1(\lambda_{n-1})))} \Delta_{n-1} - \frac{1}{2!}\frac{u''(u^{-1}(\overline{v}_1(\lambda_{n-1})))}{u'(u^{-1}(\overline{v}_1(\lambda_{n-1})))^3}\Delta_{n-1}^2  \nonumber \\
&+\frac{1}{3!}\left(-\frac{u'''(u^{-1}(\overline{v}_1(\lambda_{n-1})))}{u'(u^{-1}(\overline{v}_1(\lambda_{n-1})))^4}+ \frac{3 u''(u^{-1}(\overline{v}_1(\lambda_{n-1})))^2}{u'(u^{-1}(\overline{v}_1(\lambda_{n-1})))^5} \right)\Delta_{n-1}^3 \nonumber \\
&+ h_3(\overline{v}_1(\lambda_n(e \vert \lambda_{n-1})))\Delta_{n-1}^3,
\label{eq:Tayler-series-for-e}
\end{align}}
where $\Delta_{n-1}=(1-\beta)(e-\E[e_n])/(\beta\lambda_{n-1})$ and $h_3:\mathbb{R} \to \mathbb{R}$ is a function such that 
$\lim_{\Delta_{n-1} \to 0} h_3(\overline{v}_1(\lambda_n(e \vert \lambda_{n-1})))=0$.  
Taking the expectation of \eqref{eq:Tayler-series-for-e}, $\overline{\tau}_n-\overline{\tau}_{n-1}$ is written as{\footnotesize
\begin{align*}
\overline{\tau}_n -\overline{\tau}_{n-1} 
&=\E\left[ 
\frac{1}{2!} \left( \frac{-u''(u^{-1}(\overline{v}_1(\lambda_{n-1}))}{u'(u^{-1}(\overline{v}_1(\lambda_{n-1})))}
\right)\left(\frac{1-\beta}{\beta}\right)^2\Var[e_n]
 \right]+ o (\Delta_{n-1}^3).
\end{align*}
}
Note that since incomes follow a symmetric distribution, the third term on the right-hand side of the equation \eqref{eq:Tayler-series-for-e} will cancel out when the expectation is taken. 

Since $(-u''/u') \circ u^{-1}$ is a concave function, from Jensen's inequality we have{\footnotesize
\begin{align*}
\E\left[ -\frac{u''}{u'} \left( u^{-1}(\overline{v}_1(\lambda_{n-1}))
\right)
\right]
&=\E\left[-\frac{u''}{u'} \big( 
u^{-1}\big(
\E[\overline{v}_1(\lambda_{n}) \vert \overline{v}_1(\lambda_{n-1})]\big)
\big)
\right]\\
&\geq
\E\left[\E\left[-\frac{u''}{u'} \big( u^{-1}(\overline{v}_1(\lambda_n)) \big)\bigg\vert \overline{v}_1(\lambda_{n-1})
\right]
\right] \\
&=\E\left[-\frac{u''}{u'} \big( u^{-1}(\overline{v}_1(\lambda_n)) )\right].
\end{align*}
}
It follows that the second-order polynomial approximation of $\{\overline{\tau}_n-\overline{\tau}_{n-1}\}_{n \in \mathbb{N}}$ obtained from the third-order expansion is a decreasing sequence.  

We can now proceed analogously to the proof of the case where $\lambda_{n-1}$ must be defined in \eqref{eq:lambda'-near-e}. We change $\Delta_{n-1}$ in \eqref{eq:Tayler-series-for-e}  to {\footnotesize
\[\overline{\Delta}_{n-1}=\frac{(1-\beta)\big(\min\{e, \overline{e}_{\lambda_{n-1}} \}-\E\big[\min\{e, \overline{e}_{\lambda_{n-1}} \}\big]\big)}{\beta \lambda_{n-1}}.\] }
By the asymmetric distribution of conditional income $e \leq  \overline{e}_{\lambda_{n-1}}$, the third term on the right-hand side of \eqref{eq:Tayler-series-for-e} remains  
and $\overline{\tau}_n-\overline{\tau}_{n-1}$ is given by  {\footnotesize
\begin{align*}
\overline{\tau}_n-\overline{\tau}_{n-1}=\E\left[ 
\frac{1}{2!} 
\left( 
\frac{-u''(u^{-1}(\overline{v}_1(\lambda_{n-1})))}{u'(u^{-1}(\overline{v}_1(\lambda_{n-1})))}
\right)\left(\frac{1-\beta}{\beta}\right)^2
\Var\big[\min\{e, \overline{e}_{\lambda_{n-1}} \}\big]
 \right]+ o (\overline{\Delta}_{n-1}^2).
\end{align*}
}
By a similar argument, the second-order polynomial approximation of $\{\overline{\tau}_n-\overline{\tau}_{n-1}\}_{n \in \mathbb{N}}$ is a decreasing sequence, and the proof is complete.  
\end{proof}
We prove Proposition \ref{prop:sustainablity} below. 
\begin{proof}[Proof of Proposition \ref{prop:sustainablity}]
From Lemma \ref{lem:decreasing-sequence}, with the second-order polynomial approximation, we have 
\[\overline{\tau}_{n}-\overline{\tau}_{n-1} \leq \overline{\tau}_1 -\overline{\tau}_0 \quad \text{for }n \in \mathbb{N}. \]
Hence, the following inequality holds.
\begin{align*}
g_t(\lambda_0)&=\sum_{k=1}^t \alpha^k(\overline{\tau}_{k-1} - \overline{\tau}_k)-\overline{\tau}_0\\
&\geq \alpha \cdot \frac{1-\alpha^t}{1-\alpha}(\overline{\tau}_0-\overline{\tau}_1)-\overline{\tau}_0.
\end{align*}
The last term is positive if 
\[ -\overline{\tau}_0 \geq \frac{\alpha(1-\alpha^t)}{1-\alpha}(\overline{\tau}_1-\overline{\tau}_0)\]
and we see a sufficient condition for $g_t(\lambda_0) \geq 0$ for $t \in \mathbb{N}$ is $\alpha(\overline{\tau}_1 -\overline{\tau}_0)/(1-\alpha) \leq -\overline{\tau}_0$.
\end{proof}

\section{Proof of Corollary \ref{cor:sufficient}}\label{sec:appen-cor}

\begin{proof}[Proof of Corollary \ref{cor:sufficient}]
Let us first examine the case where $\lambda_1$ can be defined in \eqref{eq:lambda'}.
Applying Taylor theorem to $u^{-1}(\overline{v}_1(\lambda_1))$, $\alpha(\overline{\tau}_1-\overline{\tau}_0)/(1-\alpha)$ is expressed as {%\small
\begin{align}
\frac{\alpha(\overline{\tau}_1-\overline{\tau}_0)}{1-\alpha} &=
\frac{(u^{-1})''(\overline{v}_1(\lambda_0))}{2!}\left(\frac{1-\beta}{\beta \lambda_0}\right)^2\frac{\alpha}{1-\alpha}\Var[e_0] 
+ o (\Delta_0^3). \label{eq:initial_condition_left}
\end{align}
}
On the other hand, applying Taylor to $u^{-1}(\overline{v}_1(\lambda_E))$,  we have {%\small
\begin{align}
-\overline{\tau}_0 &=-u^{-1}(\overline{v}_1(\lambda_0)) + u^{-1}(\overline{v}_1(\lambda_E)) \nonumber \\
&=\frac{(u^{-1})'(\overline{v}_1(\lambda_0))}{1!}\Delta_E+\frac{(u^{-1})''(\overline{v}_1(\lambda_0))}{2!} \Delta_E^2+o(\Delta_E^2), \label{eq:initial_condition_right}
\end{align}
}
where $\Delta_E=\overline{v}_1(\lambda_E)-\overline{v}_1(\lambda_0)$. 
Since the first term on the right-hand side of \eqref{eq:initial_condition_right} is positive, 
it is sufficient to show the second term on the right-hand side of   \eqref{eq:initial_condition_right} is equal to or greater than the first term on the right-hand side of \eqref{eq:initial_condition_left}.
However, from the corollary assumption \eqref{ineq:sufficient-lambda_0}, we have 
\[\Delta_E^2 \geq \frac{(1-\beta)^2}{\lambda_0^2 \beta^2} \frac{\alpha}{1-\alpha}\Var[e_t],\]
which is the desired conclusion that the condition $\alpha(\overline{\tau}_1 -\overline{\tau}_0)/(1-\alpha) \leq -\overline{\tau}_0$  holds with the second-order polynomial approximation. 
For the case where $\lambda_1$ must be defined in \eqref{eq:lambda'-near-e}, we have 
{
\begin{align*}
\frac{\alpha(\overline{\tau}_1-\overline{\tau}_0)}{1-\alpha} &=
\frac{(u^{-1})''(\overline{v}_1(\lambda_0))}{2!}\left(\frac{1-\beta}{\beta \lambda_0}\right)^2\frac{\alpha}{1-\alpha}\Var\big[\min\{e_0,\overline{e}_{\lambda_0}\}\big] 
+ o (\overline{\Delta}_0^2). 
\end{align*}
}
Since $\Var[\min\{e_0,\overline{e}_{\lambda_0}\}] \leq \Var[e_0]$, \eqref{ineq:sufficient-lambda_0} is sufficient to be $\alpha(\overline{\tau}_1 -\overline{\tau}_0)/(1-\alpha) \leq -\overline{\tau}_0$. 
\end{proof}

\section{Proof of Proposition \ref{prop:bellman}}\label{sec:app-bellman}

\begin{proof}[Proof of Proposition \ref{prop:bellman}]
We observe first that if \eqref{ineq:sufficient-condition-less-than-S} holds, then $V$ defined in \eqref{eq:V} satisfies DP. Indeed, since by definition of $\tau^*$ we have $u(e_t + \tau^*(e_t \vert \lambda_t))=\overline{v}_1(\lambda_t)$, 
\begin{align}
V(\lambda_t,t) &=\E_t\big[ u(e_t + \tau^*(e_t \vert \lambda_t))\big] \nonumber \\
&=(1-\beta) u(e_t + \tau^*(e_t \vert \lambda_t)) + \beta \E_t \big[\overline{v}_1(\lambda_{t+1}(e_{t} \vert \lambda_{t}))\big] \nonumber \\
&=(1-\beta) u(e_t + \tau^*(e_t \vert \lambda_t)) + \beta \E_t \big[\E_{t+1}\big[\overline{v}_1(\lambda_{t+1}(e_{t} \vert \lambda_{t}))\big]\big] \nonumber \\
&=(1-\beta) u(e_t + \tau^*(e_t \vert \lambda_t)) + \beta \E_t[V(\lambda_{t+1}, t+1)]. \label{eq:V-Bellman}
\end{align}
But  $\tau^*$ is the solution to the problem
\begin{align}
\sup_{\tau \in \Lambda(\lambda_t)} 
(1-\beta) u(e_t + \tau(e_t \vert \lambda_t)) + \beta \E_t[V(\lambda_{t+1}, t+1)]. \label{eq:problem-tau}
\end{align}
Indeed, for given information in period $t$, the integrand of the objective function in \eqref{eq:problem-tau} multiplied by $\lambda_t$ is written as 
\begin{align*}
\lambda_{t}\big[ 
(1-\beta)&u(e_t + \tau(e_t \vert \lambda_t)) + \beta \overline{v}_1(\lambda_{t+1}(e_t \vert \lambda_{t}))
\big] \\
&=\lambda_t(1-\beta)u(e_t+\tau(e_t\mid\lambda_t))
+v_2(\lambda_t,e_t)
+\lambda_t\beta \overline v_1(\lambda_t)-\overline v_2(\lambda_t).
\end{align*}
Thus, up to the additive constant $\lambda_t\beta \overline v_1(\lambda_t)-\overline v_2(\lambda_t)$, the Bellman integrand is exactly the objective of problem \eqref{eq:PO}.
Hence $\tau^*$ attains the supremum. Therefore $\tau^*$ maximizes \eqref{eq:problem-tau}.

The task is now to check the terminal condition. Since $u$ is bounded from above,  
$\lim_{t \to \infty} V(\lambda_t,t) \beta^t =0$.
Hence, the sufficient terminal condition of \citet[][\,Thm.1]{wiszniewska2011terminal} is met, and we see
$V$ defined in \eqref{eq:V} is the value function of the Bellman equation in DP, and $\tau^*$ maximizes $J$ in \eqref{eq:def_of_J}.
\end{proof}

\end{appendices}
\bibliography{jecon_ref3}
\end{document}